\documentclass[a4paper, oneside, twocolumn, notitlepage, 10pt]{extarticle_ecoc}
\usepackage{ecoc}

\begin{document}
\selectlanguage{english}    


\title{Free space daylight ground-ground QKD in the near-IR}%


\author{
    Jan Tepper\textsuperscript{(1)}, Nils Hellerhoff\textsuperscript{(2)},
    Alberto Comin\textsuperscript{(1)}
}

\maketitle                  


\begin{strip}
    \begin{author_descr}

        \textsuperscript{(1)} Airbus Central Research and Technology, Taufkirchen, Germany,
        \textcolor{blue}{\uline{alberto.comin@airbus.com}}

        \textsuperscript{(2)} LMU München, Faculty of Physics (at time of the work)

    \end{author_descr}
\end{strip}

\renewcommand\footnotemark{}
\renewcommand\footnoterule{}


\begin{strip}
    \begin{ecoc_abstract}
        We report a daylight km-range free space QKD demonstration at 850nm obtaining a QBER of 1.9\% and a raw key-rate of 14 kbit/s. We used the BB84 protocol with polarisation encoding and two supporting optical beams for classical communication and clock synchronisation. \textcopyright2024 The Author(s)
    \end{ecoc_abstract}
\end{strip}


\section{Introduction}
Quantum key distribution (QKD) is a simple and effective technique to distribute encryption keys over optical links. Space based QKD, first demonstrated by the Micius satellite mission\cite{liao2017satellite}, has the potential to enable global quantum secure communication and has thus been the focus of numerous studies.\cite{pan2023free}

Despite the high level of security, the key rate which is achievable with long range free space QKD is still quite low, below one megabit per second.\cite{ecker2021strategies} Some of the limitations are intrinsic to the use of single photons, however the challenges posed by the atmospheric turbulence and the need of filtering out the environmental background photons can be technologically improved upon.\cite{pirandola2021limits} 

The atmospheric turbulence modulates the light wavefront, reducing the ability of focusing the received optical beam. This issue can be partly offset using adaptive optics, at the cost of increased system complexity. The background environmental photons are a more severe challenge. They directly increase the quantum bit error rate (QBER) and can even impede the communication when the QBER increases over a threshold set by the chosen protocol (about 10\% for the protocol BB84\cite{shor2000simple}). This is especially problematic for free space QKD and it can prevent day-light operations. One possible mitigation is choosing the QKD wavelength in the infrared region of the spectrum, where the solar background is reduced, yet the photon energy is still high enough to enable efficient single photon detection.

The telecom wavelength of 1550~nm is a natural choice, benefiting from the compatibility with the components developed for the terrestrial optical fibre networks. However, efficiently detecting single photons at 1550~nm requires using expensive superconducting nanowire detectors operating in vacuum at cryogenic conditions. For this reason, despite the benefits of using near-IR and even visible light for daylight QKD were evidenced before,\cite{abasifard2024ideal} many demonstrations including the Micius mission cited above, are done in the near infrared, where the cheaper and efficient silicon avalanche photodiodes can be used.\cite{liao2017satellite} 

In this contribution, we present preliminary results from tests on free-space daylight quantum key distribution (QKD) at 850~nm, employing the BB84 protocol with polarization encoding. The measurements were performed over link distances of up to 780~m, inside the Airbus campus near Munich. In the following paragraphs, we will describe the setup, the methods and results achieved in terms of quantum bit error rate (QBER) and key-rate.

\section{Methods}

\begin{figure}[t!]
    \centering
    \includegraphics[width=\linewidth]{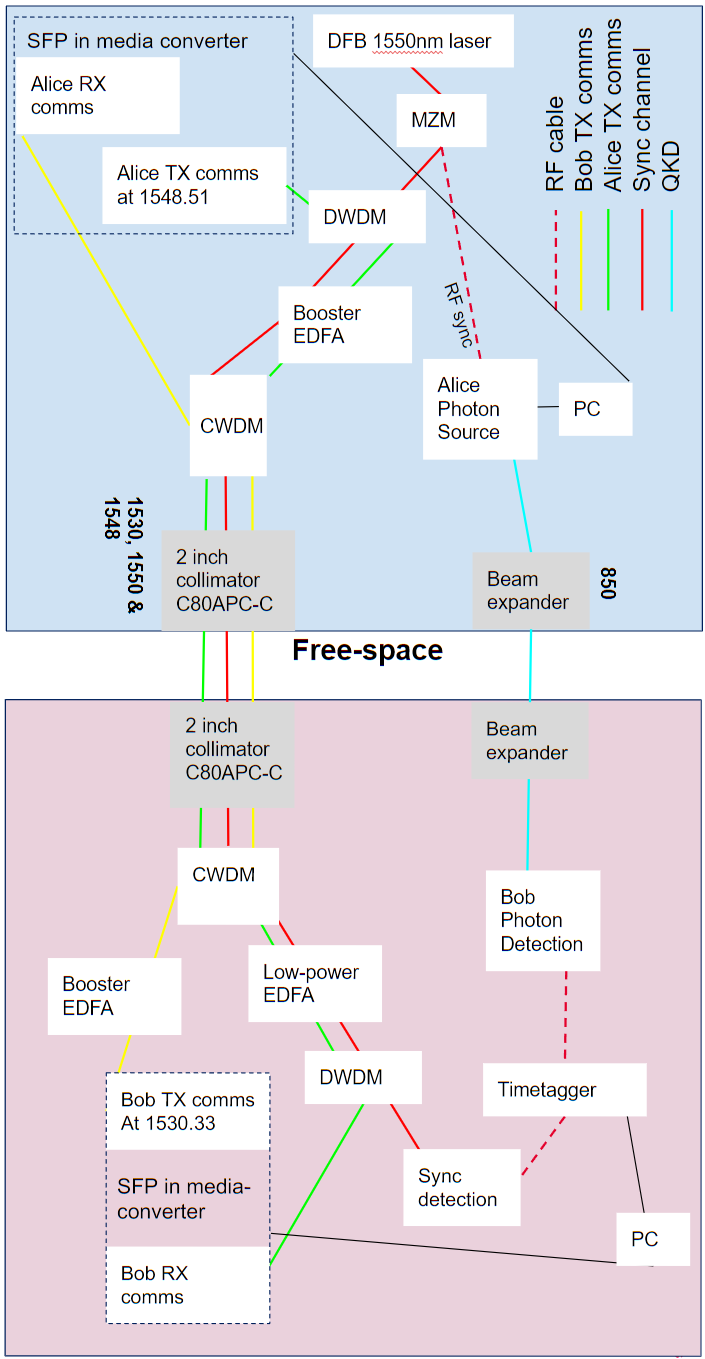}
    \caption{Block diagram of the QKD free-space system showing the sync and classical data multiplexing along the 850nm QKD signal.}
    \label{fig:figure1}
\end{figure}

The communication terminals used for the QKD experiments allowed for simultaneous classical and quantum communication. We used three free space optical channels: a quantum channel at 850~nm and two classical channels near 1550~nm. One of the classical channels was used for bidirectional (1530.33~nm \& 1548.51~nm) data transmission at 10~Gbps and the other one for clock synchronisation (1550~nm). The classical channels were amplified and multiplexed such that they could be transmitted and received through the same optics. We used a combination of spatial, temporal and spectral filtering for efficiently rejecting the environmental light. Spatial filtering was achieved by coupling the received optical beams into single mode fibers. Spectral filtering was done using commercial-off-the-shelf narrow band-pass filters. Filtering in the time domain was achieved by post-selecting the received photons based on their arrival time. The block diagram of the setup used in the second point-to-point link is shown in Fig.~\ref{fig:figure1}.

The QKD photon source was developed at the Ludwig Maximilian University of Munich (LMU) in the group of Prof. Weinfurther and it has been described in previous publications.\cite{vest2022quantum} Briefly, four vertical-cavity surface-emitting lasers (\mbox{VCSEL}) generated poissonian pulses, with a wavelength of 852~nm, pulse duration of about 200~ps and repetition rate of 100~MHz. After each VCSEL four polarizes, oriented at 0$^o$, 90$^o$, 45$^o$ and -45$^o$, encoded the qubit state in the horizontal-vertical and diagonal bases. A photonic circuit was then used to combine the output of the four VCSELs into a single mode fibre, which was coupled to free space. 

The QKD detector consisted of a small optical bench to compensate for any drift in polarisation as well as to separate the four polarisation states, four silicon avalanche single photon diodes and a high resolution time-tagger.

The software for controlling the experiments and executing the BB84 protocol was implemented with a combination of python scripts and commercial libraries in C++.

\section{Test Campaign Results}
\begin{figure}[t!]
    \centering
    \includegraphics[width=\linewidth]{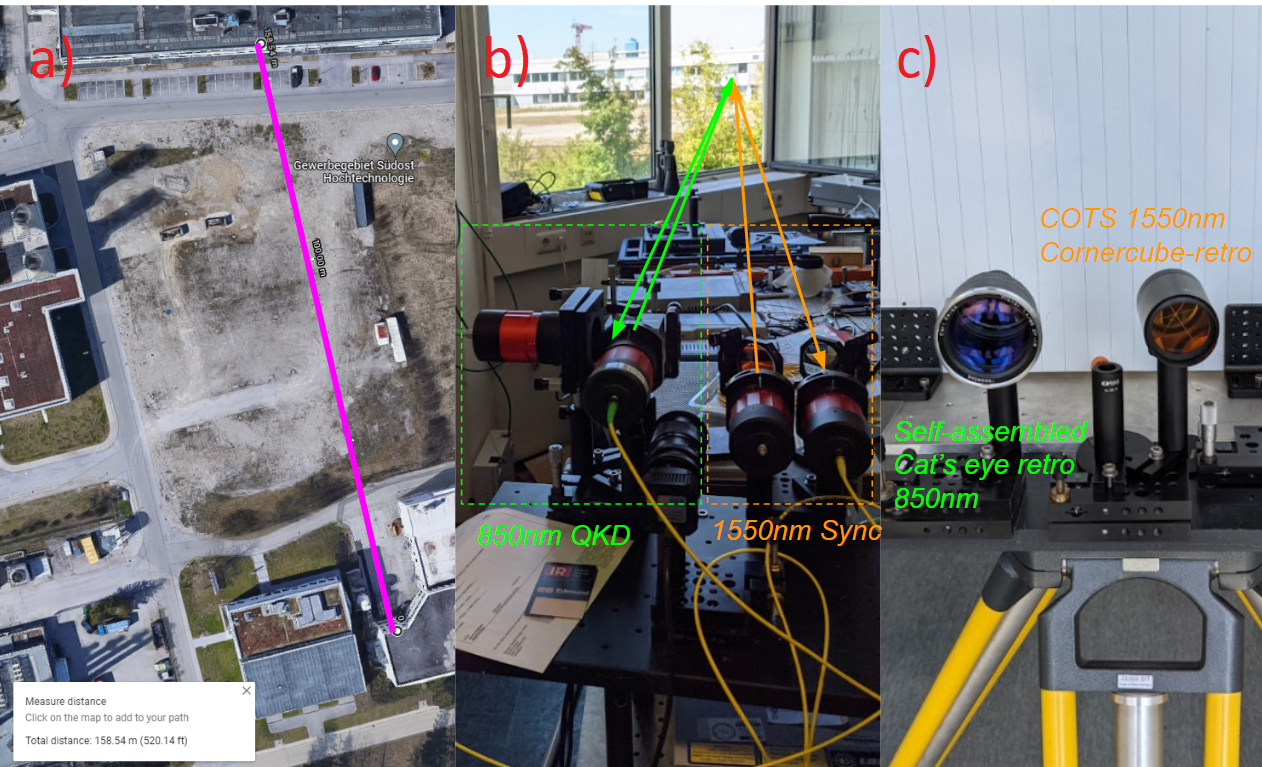}
    \caption{Optical setup for the 260~m free space QKD demonstration. a) View of the location of the experiments with the beam path highlighted in purple. b) The optics used to send and receive the quantum (green lines) and classical channel (orange lines). c) The retro-reflectors used for the quantum and the classical channels.}
    \label{fig:figure2}
\end{figure}
We set-up two free space QKD experiments with link distances of 320~m and 780~m.
The first demonstration was performed between two buildings in the Airbus campus in Taufkirchen (Germany) with the help of a retro-reflector positioned 160~m away. Fig.\ref{fig:figure2}a shows a picture of the link location with the beam path highlighted in purple. A picture taken from the transmitter side and looking towards the retro-reflectors is shown in Fig.~\ref{fig:figure2}b. The beam paths for the quantum and classical beams are highlighted with green and orange lines, respectively. 
Two different retro-reflectors, a “cat-eye” and a corner-cube were used for the quantum and classical channel, as shown in Fig.\ref{fig:figure2}c. A “cat-eye” retro-reflector was preferred for the quantum channel to minimise wavefront and polarisation distortions. It consisted of a plane mirror mounted in the focal plane of a commercial IR objective, which yielded superior optical performance compared to the COTS corner-cube type.
Different apertures were used for the transmitted and received beams. 
Since those two beams propagated along the same paths, we separated them using 50\% beam-splitters, which resulted in an additional 6~dB loss with respect to a direct link.

The experimental conditions, and the measured QBER and raw key-rate values are summarised in Tab.~\ref{tab:table1}. It should be noted that the link performance was limited by the malfunctioning of one of the VCSELs in the photon source, which significantly reduced the average photon number per pulse for the vertical polarisation.

Using a retro-reflector in an experiment results in slightly different measurement conditions with respect to a direct link, and the results might not be directly comparable. First, the retro-reflector itself may distort the propagated wavefront and, second, a potential double pass through identical turbulence cells may occur which does not occur in actual point-to-point links. 

\begin{table}[!ht]
    \centering
    \caption{QKD exchange results with the retro-reflector, for good weather and for high humidity after slight rain.} \label{tab:table1}
    \begin{tabular}{p{0.3\linewidth} | p{0.3\linewidth} | p{0.3\linewidth}}
        \hline  Entry & \textbf{Run 1} & \textbf{Run 2}                \\
        \hline  Date, time & 2022-08-18, 13:40 CET & 2022-08-18, 16:50 CET                \\
        \hline  temperature & 19$^o$C & 17$^o$C                \\
        \hline  weather & no clouds & cloudy                \\
        \hline  visibility & 10.0km & 10.0km                \\
        \hline  rain & 0~mm/h & 0.5~mm/h                \\
        \hline  wind & 4.1~m/s & 6.2~m/s                \\
        \hline  noise per APD & 4000~c/s & 700~c/s                \\
        \hline  QBER & 4.1\% & 4.9\%                \\
        \hline  raw key rate & 8.6~kbit/s & 3.4~kbit/s                \\
        \hline
    \end{tabular}
\end{table}%

\begin{figure}[t!]
    \centering
    \includegraphics[width=\linewidth]{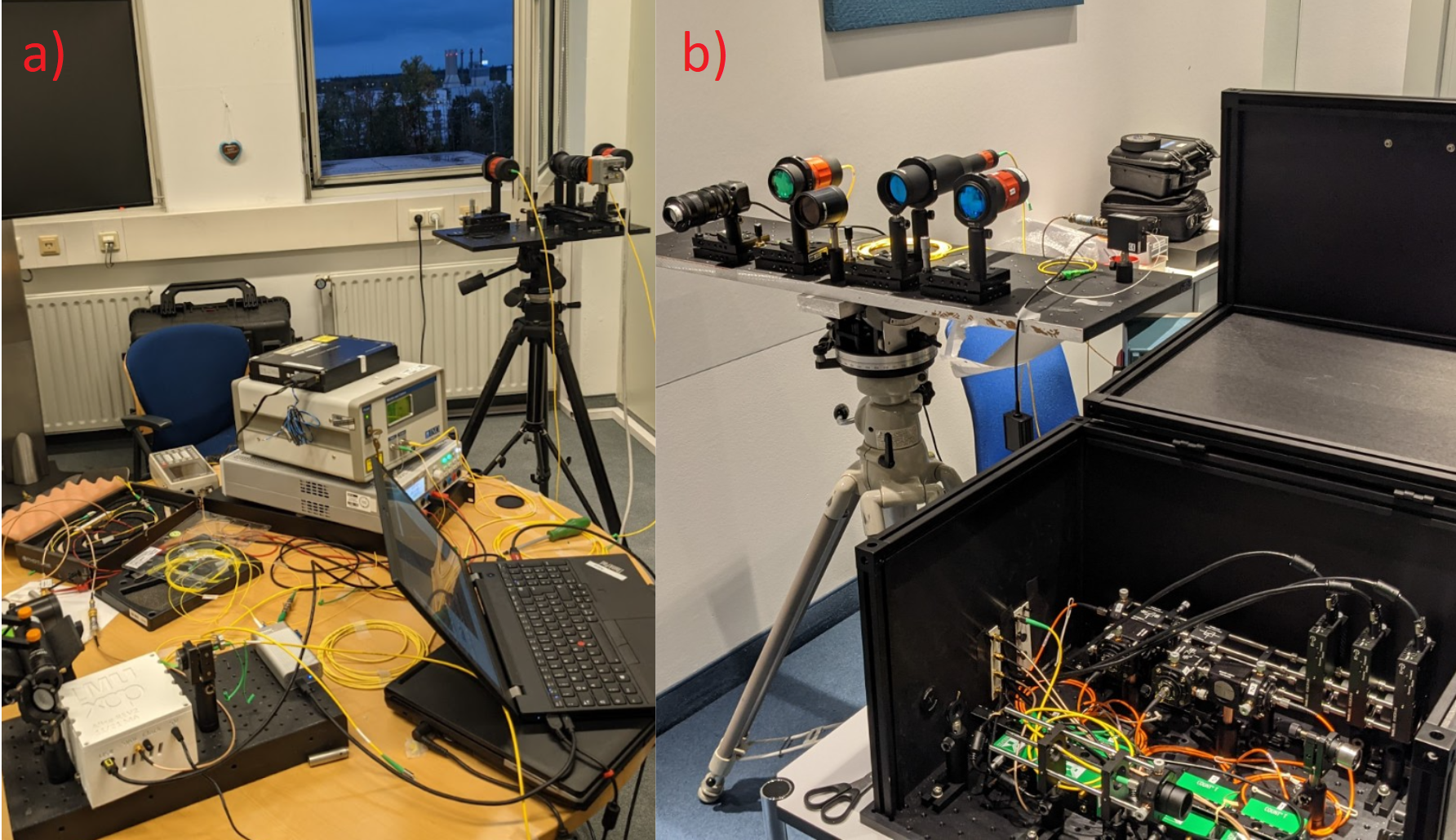}
    \caption{Optical setup for the 780m link. a) Setup in northern campus. In the front, the QKD source (white box). b) Bob setup in the southern campus. In the front, the polarisation compensation module including the four APDs (green). In the back, the receiving optics, camera and retro-reflector (for alignment only) mounted on a tripod.}
    \label{fig:figure3}
\end{figure}

For these reasons, we conducted a second test campaign, where we connected the north and south parts of the Airbus campus with a point-to-point free-space QKD link over a distance of 780~m at a height above ground of roughly 16~m. The optical setup is shown Fig.\ref{fig:figure3}. For these new experiments, the average photon number per pulse was tuned to 0.1, which was higher than the value used for the retro-reflector link, though the latter could not be known precisely due to uncertainty in the current calibration of the photon source.
In order to minimise the geometrical loss over the increased link distance, for the 850~nm quantum channel we slightly increased the beam diameter by replacing the fiber collimator used in the retro-reflector setup with a larger beam-expander. The resulting transmitted beam diameter at $1/e^2$ was 3.48cm and the maximum receivable beam diameter determined by the beam expander and the fiber's numerical aperture was 4.20cm at $1/e^2$. The latter value was determined by the clear aperture of the beam expanded and by the numerical aperture of the single mode fiber. For comparison, we also repeated the experiment with the same set of fiber collimators used in the retro-reflector link, which yielded a beam diameter at $1/e^2$ of 1.45~cm. The measurement results recorded with the two sets of optics are presented in Tab.\ref{tab:table2}: In summary we obtained a QBER of about 1.9\% and a sifted key rate of 13794~b/s.

\begin{table}[h!]
    \centering
    \caption{Results for the QKD exchange over 780m using two different sets of optics.} \label{tab:table2}
    \begin{tabular}{p{0.3\linewidth} | p{0.3\linewidth} | p{0.3\linewidth}}
        \hline  Entry & \textbf{Beam \mbox{expanders}} & \textbf{Collimators}                \\
        \hline  Date, time & 30-09-2022, 17:52CET & 30-09-2022, 18:22CET                \\
        \hline  temperature & 7$^o$C & 7$^o$C                \\
        \hline  weather & no clouds & no clouds                \\
        \hline  visibility & 2.3km & 5.0km                \\
        \hline  rain & 0~mm/h & 0~mm/h                \\
        \hline  wind & 4.1~m/s & 6.2~m/s                \\
        \hline link loss & 13~dB & 13~db \\
        \hline  noise per APD & 1500c/s & 2500c/s                \\
        \hline  QBER & 1.9~\% & 8.3~\%                \\
        \hline  raw key rate & 13.8~kbit/s & 1.4~kbit/s                \\
        \hline
    \end{tabular}
\end{table}%


\section{Conclusions}
These preliminary results confirm the feasibility of free space daylight QKD at 850~nm, across link distances in the kilometer range. They also show that utilizing a near-infrared quantum and a C-band classical channel makes it easier to spectrally isolate them and simultaneously perform classical and quantum communications. 
The seamless integration of quantum and classical free space communication, will be important for the  adoption of free space QKD technologies, and towards the development of future airborne optical networks.

\defbibnote{myprenote}{%

}
\printbibliography[prenote=myprenote]

\vspace{-4mm}

\end{document}